\documentclass[twocolumn]{aastex701}

\usepackage{lineno}
\usepackage{multirow}
\usepackage{booktabs}

\usepackage{amsmath,latexsym}    
\usepackage{graphicx}   
\usepackage{verbatim}   
\usepackage{color}      
\usepackage{subfigure}  
\usepackage{hyperref}   
\usepackage{tensor}     
\usepackage{amssymb}
\usepackage{enumerate}	
\usepackage{braket}		

\received{}
\revised{}
\accepted{}
\published{}

\shortauthors{Roy Choudhury, Okumura \& Umetsu}

\begin{document}

\title{
Cosmological constraints on nonphantom dynamical dark energy with DESI Data Release 2 Baryon Acoustic Oscillations: A 3$\sigma$+ lensing anomaly
}

\author[orcid=0000-0002-5589-3454,sname='Roy Chhoudhury']{Shouvik Roy Choudhury}
\affiliation{Institute of Astronomy and Astrophysics, Academia Sinica, No. 1, Section 4, Roosevelt Road, Taipei 106319, Taiwan}
\email[show]{sroy@asiaa.sinica.edu.tw}

\author[orcid=0000-0002-8942-9772,sname='Okumura']{Teppei Okumura}
\affiliation{Institute of Astronomy and Astrophysics, Academia Sinica, No. 1, Section 4, Roosevelt Road, Taipei 106319, Taiwan}
\affiliation{Kavli Institute for the Physics and Mathematics of the Universe (WPI), UTIAS, The University of Tokyo, Chiba 277-8583, Japan}
\email{tokumura@asiaa.sinica.edu.tw}

\author[orcid=0000-0002-7196-4822,sname='Umetsu']{Keiichi Umetsu}
\affiliation{Institute of Astronomy and Astrophysics, Academia Sinica, No. 1, Section 4, Roosevelt Road, Taipei 106319, Taiwan}
\email{keiichi@asiaa.sinica.edu.tw}

\begin{abstract}
We consider a 12-parameter cosmological model with non-phantom dynamical dark energy (NPDDE), where non-phantom implies that the equation of state (EoS) of dark energy (DE), $w(z)\geq-1$ for all redshifts $z$. Thus, the DE-EoS covers the parameter space corresponding to the popular single scalar-field dark energy models, i.e., Quintessence. The cosmological model comprises 6 parameters of the $\Lambda$-Cold Dark Matter ($\Lambda$CDM) model, and additionally the dynamical DE EoS parameters ($w_0$, $w_a$), the scaling of the lensing amplitude ($A_{\rm lens}$), sum of the neutrino masses ($\sum m_\nu$), the effective number of non-photon relativistic degrees of freedom ($N_{\rm eff}$), and the running of the scalar spectral index ($\alpha_s$). We derive constraints on the parameters by combining the latest Dark Energy Spectroscopic Instrument (DESI) Data Release (DR) 2 Baryon Acoustic Oscillation (BAO) measurements with cosmic microwave background (CMB) power spectra from Planck Public Release (PR) 4, CMB lensing data from Planck PR4 and Atacama Cosmology Telescope (ACT) DR6, uncalibrated Type Ia supernovae (SNe) data from the Pantheon+ and Dark Energy Survey (DES) Year 5 (DESY5) samples, and Weak Lensing (WL) data from DES Year 1. Our major finding is that with CMB+BAO+WL and CMB+BAO+SNe+WL, we find 3$\sigma$+ evidence for $A_{\rm lens} >1$ even with Planck PR4, indicating a higher than expected CMB lensing amplitude relative to the NPDDE prediction of unity. This implies that for cosmology to accommodate realistic quintessence-like dark energy models (as opposed to unrealistic phantom DE), one would also need to explain a relatively significant presence of the lensing anomaly.
\end{abstract}




\section{Introduction}
\label{sec:1}
The nature of dark energy (DE) remains one of the central open questions in modern cosmology. The $\Lambda$-Cold Dark Matter ($\Lambda$CDM) model has achieved remarkable success in accounting for a wide array of cosmological observations across both low and high redshifts. However, recent results from the Dark Energy Spectroscopic Instrument (DESI) collaboration \cite{DESI:2024mwx,DESI:2024hhd,DESI:2025zgx,DESI:2025fii} have provided intriguing hints for an evolving dark energy component, including a possible phantom crossing around redshift $z \simeq 0.5$ when analyzed with several commonly used parameterizations of dynamical dark energy \cite{DESI:2025fii}. When combined with Cosmic Microwave Background (CMB) and Type Ia supernovae (SNe~Ia) data, the DESI Data Release (DR) 2 measurements of Baryon Acoustic Oscillations (BAO) disfavor the cosmological constant at the level of $2.8\sigma$, $3.8\sigma$, and $4.2\sigma$, depending on the supernovae (SNe) compilation employed---PantheonPlus \cite{Brout:2022vxf}, Union3 \cite{Rubin:2023ovl}, and Dark Energy Survey (DES) Year 5 (DESY5) \cite{DES:2024tys}, respectively. These results were obtained within an eight-parameter cosmological model using the Chevallier-Polarski-Linder (CPL) parameterization \cite{Chevallier:2000qy,Linder:2002et} of the dark energy equation of state (EoS), defined as 
\begin{equation}\label{eqn1}
    w(z) \equiv w_{0} + w_{a}\,\frac{z}{1+z},
\end{equation}
where $z$ is the redshift. The potentially far-reaching implications of these findings have triggered an extensive wave of follow-up investigations into the physics of dark energy \cite[see e.g.,][]{RoyChoudhury:2024wri, RoyChoudhury:2025dhe,Tada:2024znt,Berghaus:2024kra,Park:2024vrw,Yin:2024hba,Shlivko:2024llw,Cortes:2024lgw,DESI:2024kob,Carloni:2024zpl,Croker:2024jfg,Mukherjee:2024ryz,Roy:2024kni,Wang:2024dka,Gialamas:2024lyw,Orchard:2024bve,Giare:2024gpk,Dinda:2024ktd,Jiang:2024xnu,Reboucas:2024smm,Bhattacharya:2024hep,Pang:2024qyh,Ramadan:2024kmn,Wolf:2024eph,Efstathiou:2024xcq,Kessler:2025kju,Gao:2025ozb,Borghetto:2025jrk,Wolf:2025jlc,Huang:2025som,Peng:2025nez,Tsedrik:2025cwc,Reeves:2025axp,Yang:2025kgc,Park:2025azv,Odintsov:2024woi,Shim:2024mrl,Gao:2024ily,Payeur:2024dnq,Ye:2024zpk,Luongo:2024zhc,Wolf:2024stt,Giare:2024oil,Chakraborty:2025syu,Elbers:2024sha,Giare:2025pzu,Wolf:2025dss,Ye:2025ulq,Colgain:2024xqj,Colgain:2025nzf,Ye:2024ywg,Silva:2025hxw,Park:2024pew,Luciano:2025ykr,Fazzari:2025lzd,Zhou:2025nkb,Wu:2025vfs,vanderWesthuizen:2025mnw,vanderWesthuizen:2025rip,vanderWesthuizen:2025vcb,Chen:2025ywv,Barua:2025adv,Chaudhary:2025vzy,Wang:2025znm,Feng:2025wbz,Braglia:2025gdo,Li:2025htp,Silva:2025twg,Lee:2025pzo,Li:2025dwz,Mishra:2025goj,Mazumdar:2025smh,Liu:2025myr,Hogas:2025ahb,Giare:2025ath,Gialamas:2025pwv,Ozulker:2025ehg,Wang:2025dtk,Luciano:2025elo,vanderWesthuizen:2025iam,Mukherjee:2025ytj,Moffat:2025jmx,Bayat:2025xfr,Giani:2025hhs,Cheng:2025lod,Ye:2025ark,Jhaveri:2025neg,Scherer:2025esj,Dinda:2025iaq,Mirpoorian:2025rfp,Paliathanasis:2025cuc,Kou:2025yfr,Dinda:2025hiu,Wolf:2025acj,GarciaEscudero:2025lef,Toomey:2025xyo,Adam:2025kve,Huang:2025xyf,Sanyal:2025udg,Plaza:2025nip,Yang:2025oax,Wang:2025dzn,Yang:2025uyv,Petri:2025swg,Meetei:2025mhu,Roy:2025cxk,Arora:2025msq,Akita:2025txo,Ishak:2025cay,Zapata:2025ngr,Chaudhary:2025pcc,Shah:2025vnt,Guedezounme:2025wav,Goldstein:2025epp,Qiang:2025cxp,Sharma:2025qmv,Nojiri:2025low,Li:2025ops,Herold:2025hkb,An:2025vfz,Lee:2025hjw,Wang:2025vfb,Bhattacharya:2024kxp,Li:2024qso,Li:2025owk,Chaudhary:2025uzr,Chakraborty:2024xas,Hossain:2025grx}.

The line at $w = -1$ (corresponsing to the cosmological constant or $\Lambda$) separates the behavior of dark energy into two regimes: phantom ($w < -1$) and non-phantom ($w \geq -1$). In this work, we focus only on the non-phantom scenario and exclude the phantom region. From a quantum field theory standpoint, models with a single scalar field cannot cross the $w = -1$ boundary, while more general models that do allow it require additional degrees of freedom to maintain gravitational stability. Theories that accommodate phantom dark energy typically suffer from issues such as Lorentz violation, unstable vacuum, superluminal modes, ghosts, non-locality, or instability to quantum corrections \citep{Caldwell:2003vq,Nesseris:2004uj,Vikman:2004dc,Fang:2008sn}. On the other hand, single scalar field models like quintessence are theoretically well motivated, avoid these problems, and remain strictly non-phantom \citep{Linder:2007wa,Caldwell:2005tm,Zlatev:1998tr,Wang:2004py}. Therefore, in this work we restrict our analysis to the non-phantom dynamics throughout the evolution of the universe, i.e., w(z) $\geq$-1 for all redshifts $z$. We, however, emphasize here, that an apparent effective equation of state $w < -1$ can arise without the above mentioned pathologies in more complicated but stable frameworks, such as multi-field dark energy models \citep{Feng:2004ad,Cai:2009zp,Guo:2004fq,Akrami:2020zfz,Eskilt:2022zky} or in scenarios involving interactions between dark energy and dark matter \citep{Amendola:1999er,Farrar:2003uw,vandeBruck:2016hpz}.

It should be noted that in the 8-parameter $w_0w_a$CDM model analyzed by the DESI Collaboration, the non-phantom region of dynamical dark energy is excluded at more than $2\sigma$ significance when using CMB+BAO or CMB+BAO+SNe dataset combinations. However, two recent works \cite{RoyChoudhury:2024wri, RoyChoudhury:2025dhe} showed that extending the dynamical dark energy framework to a 12-parameter model allows a substantial portion of the non-phantom region to remain consistent within $2\sigma$ when using CMB+BAO+Pantheon+. Moreover, in this extended parameter space, the non-phantom region is not conclusively ruled out beyond $2\sigma$ by either the CMB+BAO or the CMB+BAO+DESY5 dataset combinations. Therefore, analyzing the non-phantom dynamical dark energy model (hereafter NPDDE) is well motivated in the 12-parameter setup considered in \cite{RoyChoudhury:2024wri, RoyChoudhury:2025dhe}, though it is disfavored in the minimal 8-parameter $w_0w_a$CDM model studied by the DESI Collaboration. 

The extended model includes the six baseline $\Lambda$CDM parameters along with the following extensions: the dynamical dark energy equation of state parameters ($w_0$, $w_a$) in the CPL parametrization, the sum of neutrino masses ($\sum m_{\nu}$), the effective number of relativistic species ($N_{\rm eff}$), the lensing amplitude scaling parameter ($A_{\rm lens}$), and the running of the scalar spectral index ($\alpha_s$)\footnote{For previous studies in such largely extended parameter spaces, see \cite{DiValentino:2015ola,DiValentino:2016hlg,DiValentino:2017zyq,Poulin:2018zxs,RoyChoudhury:2018vnm,DiValentino:2019dzu}.}. For the CMB data, we employ the latest Planck Public Release 4 (PR4, 2020) likelihoods, namely HiLLiPoP and LoLLiPoP \citep{Tristram:2023haj}, together with Planck PR4 lensing combined with ACT DR6 lensing likelihoods \citep{ACT:2023kun}. For BAO, we use the DESI DR2 BAO likelihoods \citep{DESI:2025zgx}, while for supernovae we adopt the most recent uncalibrated Type Ia Supernova likelihoods: Pantheon+ \citep{Brout:2022vxf} and DESY5 \citep{DES:2024tys}. In addition, we make use of the Dark Energy Survey Year 1 (DESY1) galaxy clustering and weak lensing data \citep{DES:2017myr}\footnote{We note that more recent weak lensing data from DES Year 3 are available; however, the corresponding likelihoods have been released only for \texttt{CosmoSIS} \citep{Zuntz:2014csq} and not for \texttt{Cobaya} \citep{Torrado:2020dgo}, which is the MCMC code adopted in this work.}.

We now outline the motivation for extending the $w_0w_a$CDM model (apart from allowing for the NPDDE parameter space). Neutrinos, while massless in the Standard Model, are known from oscillation experiments to be massive, requiring at least two non-zero masses with mass-squared splittings $\Delta m_{21}^2 \simeq 7.42 \times 10^{-5}$ eV$^2$ and $|\Delta m_{31}^2| \simeq 2.51 \times 10^{-3}$ eV$^2$. This implies two possible mass orderings: Normal ($\sum m_{\nu} > 0.057$ eV) and Inverted ($\sum m_{\nu} > 0.096$ eV) \citep{Esteban:2020cvm}. Hence, the inclusion of $\sum m_{\nu}$ represents a simple and well-motivated extension to the baseline model. Dynamical dark energy, parameterized through the CPL form ($w_0$, $w_a$), is also relevant since it introduces geometric degeneracies with $\sum m_\nu$ \citep{RoyChoudhury:2019hls}. Notably, the DESI Collaboration has reported that the preferred $w_0$–$w_a$ region lies in the phantom regime \citep{DESI:2024mwx,DESI:2024hhd,DESI:2025zgx}, which relaxes the bounds on $\sum m_{\nu}$ by more than a factor of two due to the well-known degeneracy between the dark energy equation of state and the neutrino mass \citep{Hannestad:2005gj}. This degeneracy is of significant importance here, as the NPDDE parameter space ($w(z)\geq-1$) prefers lower values of $\sum m_{\nu}$ than a cosmological constant ($w=-1$), and thus stronger bounds on $\sum m_{\nu}$ are achieved in a cosmological model with NPDDE than $\Lambda$, a fact first noticed in 2018 by \cite{Vagnozzi:2018jhn,RoyChoudhury:2018vnm,RoyChoudhury:2018gay}. Due to this degeneracy, it is expected that the bounds on $\sum m_{\nu}$ obtained in this paper will be stronger than our recent works \cite{RoyChoudhury:2024wri,RoyChoudhury:2025dhe},\footnote{For earlier bounds on $\sum m_{\nu}$ in the literature, see e.g., \cite{RoyChoudhury:2018gay,RoyChoudhury:2019hls,Tanseri:2022zfe,Vagnozzi:2017ovm,Vagnozzi:2018jhn,RoyChoudhury:2018vnm,DiValentino:2021hoh,Giusarma:2018jei,Reeves:2022aoi,Yang:2017amu,RoyChoudhury:2018bsd}. For more recent studies, see \cite{Shao:2024mag,Jiang:2024viw, Herold:2024nvk, RoyChoudhury:2024wri,Bertolez-Martinez:2024wez,Wang:2025ker,RoyChoudhury:2025dhe}}  where the full dynamical DE parameter space was allowed. 

Massive neutrinos, upon becoming non-relativistic, suppress the growth of structure on small scales, thereby affecting CMB lensing and enhancing small-scale anisotropy correlations \citep{RoyChoudhury:2020das, Lesgourgues:2012uu}. The $>2\sigma$ lensing anomaly in the Planck PR3 (2018) likelihoods ($A_{\rm lens} = 1.18 \pm 0.065$ at 68\%) \citep{Planck:2018nkj, Planck:2018vyg} has essentially disappeared ($<1\sigma$) in Planck PR4 \citep{Tristram:2023haj} in the $\Lambda$CDM model, which motivates the use of PR4 for robust neutrino mass constraints\footnote{The $A_{\rm lens}$ parameter was first defined in \citep{Calabrese:2008rt}.}. Since $\sum m_{\nu}$ and $A_{\rm lens}$ are correlated, including $A_{\rm lens}$ as a free parameter in combination with Planck PR4 likelihoods helps mitigate potential systematic biases in neutrino mass inference.

We further note that recent studies have identified a correlation between $A_{\rm lens}$ and the dark energy equation of state parameters \citep{Chan-GyungPark:2025cri,Park:2024pew}, which leads to a relaxation of the constraints on $w_0$ and $w_a$. A similar relaxation was observed in our earlier works \citep{RoyChoudhury:2024wri, RoyChoudhury:2025dhe}. This correlation effectively reduces the apparent tension with $\Lambda$CDM (and NPDDE) relative to that reported by the DESI Collaboration \citep{DESI:2025zgx}. However, this reduction in tension comes at the cost of increasing the preferred value of $A_{\rm lens}$, thereby shifting it away from consistency with the canonical value $A_{\rm lens} = 1$. It will be, thus, of great importance to analyse what happens to the $A_{\rm lens}$ parameter (and thus the lensing anomaly) when the parameter space is strictly restricted to the NPDDE region that corresponds to realistic quantum field theories of dark energy. Weak Lensing data can further exacerbate the lensing anomaly, as we noticed in our recent work \cite{RoyChoudhury:2025dhe}, where we showed for the first time that even with Planck PR4 likelihoods, addition of a non-CMB data like DES Year 1 Weak Lensing with CMB+BAO+SNe can lead to a $>$2$\sigma$ lensing anomaly in this 12 parameter model, but with full range of $w_0$, $w_a$ allowed. 

\begin{figure}[tbp]
	\begin{center}
		\includegraphics[width=.99\linewidth]{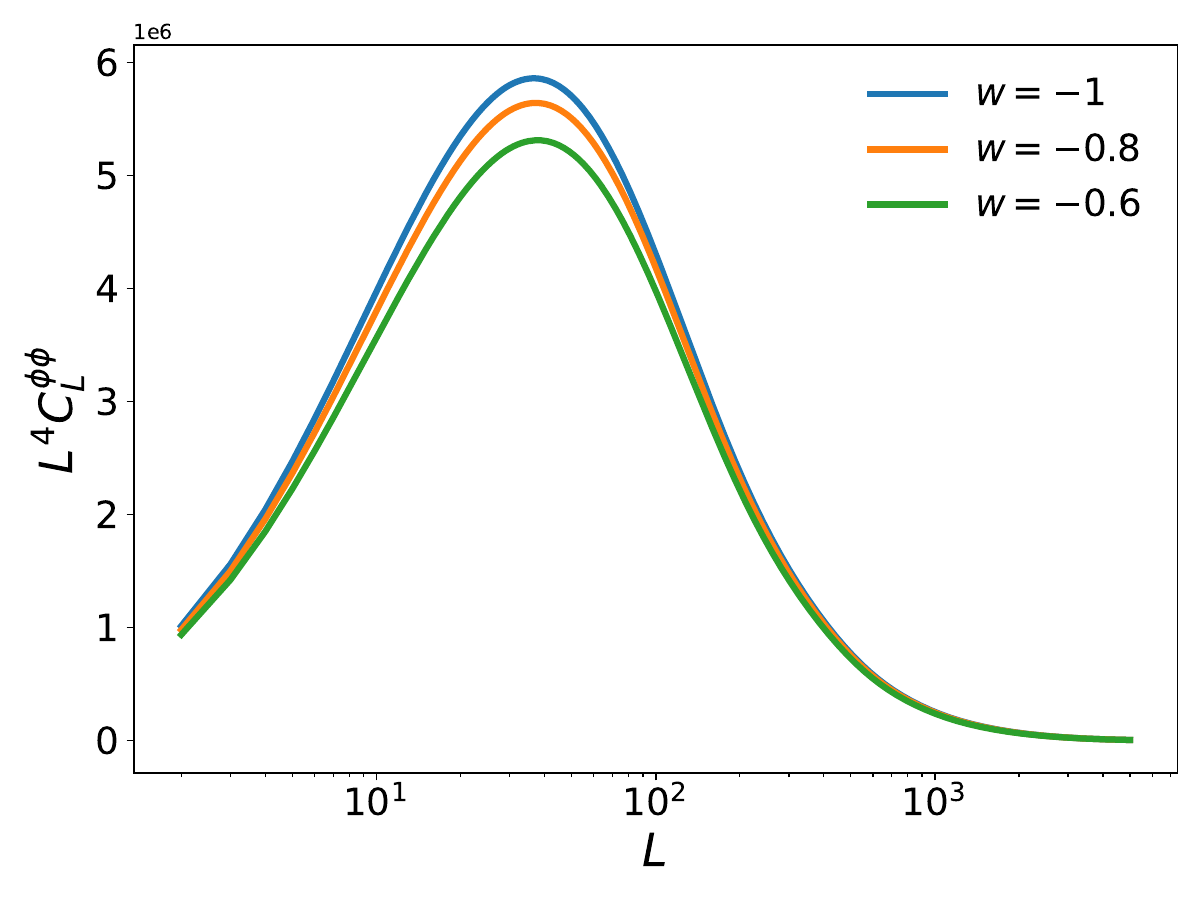}
		\caption{\label{fig:0} The CMB lensing-potential powerspectrum for different values of the constant DE EoS $w$. The angular scale of sound horizon at recombination, $\theta_s^*$, is kept fixed by adjusting $H_0$.}
	\end{center}
\end{figure}

The redshift at which the matter and dark energy densities become equal, denoted by \(z_{\rm eq}\), is obtained from the condition \(\rho_m(z_{\rm eq}) = \rho_{\rm DE}(z_{\rm eq})\). 
Assuming a constant dark energy equation of state \(w\), the matter and dark energy densities evolve as 
\(\rho_m(z) = \rho_{m0}(1+z)^3\) and \(\rho_{\rm DE}(z) = \rho_{\rm DE,0}(1+z)^{3(1+w)}\), 
where \(\rho_{m0}\) and \(\rho_{\rm DE,0}\) are their present-day values. 
This leads to
\begin{equation}
	1 + z_{\rm eq} = 
	\left( \frac{\Omega_{\rm DE,0}}{\Omega_{m0}} \right)^{-1/(3w)} ,
	\label{eq:zeq}
\end{equation}
where \(\Omega_{m0}\) and \(\Omega_{\rm DE,0}\) are the present-day fractional density parameters of matter and dark energy, respectively. 
For \(w = -1\) (the $\Lambda$CDM case), Eq.~(\ref{eq:zeq}) reduces to \(1 + z_{\rm eq} = (\Omega_{\rm DE,0}/\Omega_{m0})^{1/3}\). 
Since the exponent \(-1/(3w)\) increases as \(w\) becomes less negative, non-phantom dark energy models (\(w > -1\)) yield a higher \(z_{\rm eq}\), implying that matter–dark energy equality occurs earlier in cosmic history compared to $\Lambda$CDM, for the same value of \(\Omega_{m0}\).
In contrast, phantom models (\(w < -1\)) result in a lower \(z_{\rm eq}\), meaning that equality occurs later. 
This earlier onset of dark energy dominance in non-phantom models suppresses the growth of large-scale structure and reduces the amount of CMB lensing relative to $\Lambda$CDM. We can visualize this effect in Figure~\ref{fig:0}, where we have plotted the CMB lensing-potential power spectrum $C_L^{\phi\phi}$ for different values of the constant DE EoS $w$. The deeper we go into the non-phantom regime larger the suppression of the power spectrum. Thus, a correlation between the DE EoS and $A_{\rm lens}$ is naturally expected and a preference for $A_{\rm lens}>1$ essentially indicates that the data prefers more CMB lensing than what is theoretically predicted by the non-phantom model. 


The scaling of the lensing amplitude $A_{\rm lens}$ is also degenerate with the spatial curvature parameter $\Omega_k$ \citep{DiValentino:2019qzk}, linking the lensing anomaly to the preference for a nonzero curvature \citep{DiValentino:2019qzk, Handley:2019tkm}. In particular, Planck PR3 (2018) data alone favored a closed universe ($\Omega_k < 0$), excluding $\Omega_k=0$ at more than $2\sigma$. This preference for a closed universe, however, vanishes once BAO and supernova datasets are included \citep{Planck:2018vyg}. By contrast, the lensing anomaly persisted above $2\sigma$ even when these additional datasets were combined with Planck PR3. The DESI Collaboration has likewise reported no significant deviation from flatness in dynamical dark energy models \citep{DESI:2025zgx}, albeit in the $\Lambda$CDM+$\Omega_k$ model a 2$\sigma$+ preference for positive $\Omega_k$ is seen \cite{DESI:2025zgx,Chen:2025mlf}. Separate studies using physical models for quintessence like exponential quintessence \cite{Akrami:2025zlb} or thawing quintessence \cite{Dinda:2025iaq} also show a $2\sigma$+ preference for a nonzero positive curvature. However, these last two studies only make use of background data, neglecting the perturbation information encoded in the full CMB measurements. Nevertheless, it may be worthwhile to investigate the role of spatial curvature by allowing the cosmological model to vary in $\Omega_k$. Since MCMC analyses with nonzero $\Omega_k$ are computationally expensive, we defer such an exploration to future work. In the present study, we concentrate on the lensing anomaly and therefore fix $\Omega_k = 0$ throughout our analysis, i.e., we restrict ourselves to a spatially flat universe.

Another key motivation arises from the Hubble tension, a $\sim 4.6\sigma$ discrepancy between the local SH0ES measurement ($H_0 = 73.04 \pm 1.04$ km/s/Mpc) \citep{Riess:2021jrx} and the Planck PR4 inference ($H_0 = 67.64 \pm 0.52$ km/s/Mpc) \citep{Tristram:2023haj} within $\Lambda$CDM. This tension may be alleviated by phantom dark energy \citep{RoyChoudhury:2019hls, Vagnozzi:2019ezj} or by a fully thermalized additional radiation species ($\Delta N_{\rm eff} \sim 1$), although the latter is disfavored by data \citep{Planck:2018vyg, Vagnozzi:2023nrq, RoyChoudhury:2020dmd, Choudhury:2021dsc, RoyChoudhury:2022rva, Bostan:2023ped}. We nevertheless include $N_{\rm eff}$ as a free parameter to test the extent to which the $H_0$ tension can be reduced in a 12-parameter framework, even if we do not expect to fully resolve it. A larger $N_{\rm eff}$ increases the pre-recombination expansion rate, which in turn raises the inferred value of $H_0$. Given the significance of the Hubble tension in contemporary cosmology, we consider varying $N_{\rm eff}$ an essential extension.

Finally, we include the running of the scalar spectral index, $\alpha_s$. In slow-roll inflationary models, one typically expects $\log_{10}|\alpha_s| \sim -3.2$ \citep{Garcia-Bellido:2014gna}, though alternative inflationary scenarios can produce significantly larger deviations (see, e.g., \cite{Easther:2006tv,Kohri:2014jma,Chung:2003iu}). Allowing $\alpha_s$ to vary enables us to test for such possible departures from the standard slow-roll expectation.

Beyond our general motivation to constrain the parameters of a cosmological model with NPDDE, the primary goals of this paper are threefold:  
\begin{enumerate}  
	\item To further investigate the lensing anomaly (or $A_{\rm lens}$ anomaly) in the 12-parameter NPDDE framework, particularly in the presence of weak lensing data.  
	\item To re-assess the bounds on $\sum m_{\nu}$ in an NPDDE model using the new DESI DR2 BAO likelihoods.  
	\item To examine the robustness of the Hubble tension \citep{Riess:2021jrx,Knox:2019rjx,Banerjee:2020xcn,Lee:2022cyh} in this significantly extended parameter space.  
\end{enumerate}  
With the release of DESI DR2 BAO data, we consider this an opportune moment to revisit and update the constraints in an extended cosmological model with non-phantom dark energy. The resulting constraints are expected to be of considerable relevance to both the cosmology and particle physics communities.  

The remainder of the paper is organized as follows. Section~\ref{sec:2} describes the analysis methodology. Section~\ref{sec:3} presents and discusses the results of our statistical analysis. Finally, Section~\ref{sec:4} summarizes our conclusions.


\section{Analysis methodology} \label{sec:2}
In Section~\ref{sec:2.1}, we describe the cosmological model, the parameter sampling and plotting codes, and the priors adopted for the parameters. Section~\ref{sec:2.2} provides an overview of the cosmological datasets employed in this work.

\subsection{Cosmological model and parameter sampling}\label{sec:2.1}
Here is the parameter vector for this extended model with 12 parameters :
\begin{align}
\theta \equiv &\left[\omega_c, ~\omega_b, ~\Theta_s^{*},~\tau, ~n_s, ~{\rm{ln}}(10^{10} A_s), w_{0, \rm DE}, w_{a, \rm DE}, \right. \nonumber \\
&\qquad \qquad \left. N_{\textrm{eff}}, \sum m_{\nu}, \alpha_s, A_{\textrm{lens}} \right].
\label{eqn:3}
\end{align}
%
The first six parameters correspond to the standard $\Lambda$CDM model: the present-day cold dark matter energy density, $\omega_c \equiv \Omega_c h^2$; the present-day baryon energy density, $\omega_b \equiv \Omega_b h^2$; the reionization optical depth, $\tau$; the scalar spectral index, $n_s$; and the amplitude of the primordial scalar power spectrum, $A_s$ (both evaluated at the pivot scale $k_* = 0.05~\mathrm{Mpc}^{-1}$). Additionally, $\Theta_s^{*}$ denotes the ratio of the sound horizon to the angular diameter distance at the epoch of photon decoupling.

The remaining six parameters extend the $\Lambda$CDM framework. For the CPL parametrization of the dark energy equation of state, we use the notation $(w_{0,\rm DE}, w_{a,\rm DE})$ interchangeably with $(w_0, w_a)$. The other extended parameters, as discussed in the Introduction, include the effective number of non-photon relativistic species ($N_{\rm eff}$), the sum of neutrino masses ($\sum m_{\nu}$), the running of the scalar spectral index ($\alpha_s$), and the lensing amplitude scaling parameter ($A_{\rm lens}$).

Here, given equation (\ref{eqn1}), $w(z=0) = w_{0,\rm DE}$ denotes the present-day dark energy equation of state (EoS), while $w(z \to \infty) = w_{0,\rm DE} + w_{a,\rm DE}$ corresponds to its value in the early universe. The function $w(z)$ evolves monotonically between these two limits. Consequently, in order to restrict the analysis to the NPDDE region of parameter space, i.e., $w(z) \geq -1$, it suffices to impose the hard priors  
\begin{equation}  
	w_{0,\rm DE} \geq -1, \qquad w_{0,\rm DE} + w_{a,\rm DE} \geq -1.  
\end{equation}  

We adopt a degenerate hierarchy for neutrino masses, where all three neutrinos have equal masses, $m_i = \sum m_{\nu}/3$ for $i = 1,2,3$, and impose a prior $\sum m_{\nu} \geq 0$. This choice is well motivated, as cosmological observations primarily constrain the total neutrino mass through its impact on the energy density \cite{Lesgourgues:2012uu}, and even future cosmological data are expected to remain insensitive to the small mass splittings \cite{Archidiacono:2020dvx}. Moreover, forecasts indicate that assuming a degenerate hierarchy introduces only a negligible bias in the event of a detection of $\sum m_{\nu}$ \cite{Archidiacono:2020dvx}. Currently, no compelling evidence favors a particular neutrino mass hierarchy, even when combining cosmological constraints with terrestrial experiments such as neutrino oscillation and beta decay measurements \cite{Gariazzo:2022ahe}.

Since we allow for a nonzero running of the scalar spectral index, $\alpha_s \equiv d n_s / d\ln k$, with $k$ the wave number, we adopt a standard running power-law model for the primordial scalar power spectrum, given by

\begin{equation}
\ln \mathcal{P}_s(k) = \ln A_s + (n_s - 1) \ln\left(\frac{k}{k_*}\right) + \frac{\alpha_s}{2} \left[ \ln\left(\frac{k}{k_*}\right) \right]^2.
\end{equation}

Small values of $\log_{10}|\alpha_s| \simeq -3.2$ naturally arise in slow-roll inflationary models \cite{Garcia-Bellido:2014gna}, although certain inflationary scenarios can produce larger values (see, e.g., \cite{Easther:2006tv,Kohri:2014jma,Chung:2003iu}).

\textbf{Parameter Sampling}: For all Markov Chain Monte Carlo (MCMC) analyses presented in this work, we employ the cosmological inference framework \texttt{Cobaya} \cite{Torrado:2020dgo,2019ascl.soft10019T}, with theoretical predictions computed using the Boltzmann solver \texttt{CAMB} \cite{Lewis:1999bs,Howlett:2012mh}. When including the combined Planck PR4 + ACT DR6 lensing likelihood, we adopt the high-precision settings recommended by the ACT collaboration.

Chain convergence is assessed using the Gelman-Rubin statistics \cite{doi:10.1080/10618600.1998.10474787}, ensuring that all chains satisfy the criterion $R-1<0.01$. Parameter constraints and the figures shown in this paper are obtained using \texttt{GetDist} \cite{Lewis:2019xzd}. Broad, uniform priors are imposed on all cosmological parameters, as summarized in Table~\ref{table:1}.

\begin{table}
	\begin{center}
		\begin{tabular}{c c}
			\hline
			Parameter                    & Prior\\
			\hline
			$\Omega_{\rm b} h^2$         & [0.005, 0.1]\\
			$\Omega_{\rm c} h^2$         & [0.001, 0.99]\\
			$\tau$                       & [0.01, 0.8]\\
			$n_s$                        & [0.8, 1.2]\\
			${\rm{ln}}(10^{10} A_s)$         & [1.61, 3.91]\\
			$\Theta_s^{*}$             & [0.5, 10]\\ 
			$w_{0,\rm DE}$                        & [-1, -0.33]\\
			$w_{a,\rm DE}$                        & [-2, 2]\\
			$N_{\rm eff}$                & [2, 5]\\ 
			$\sum m_\nu$ (eV)            & [0, 5]\\
			$\alpha_s$                    & [-0.1, 0.1]\\
			$A_{\textrm{lens}}$          & [0.1, 2]\\
			\hline
		\end{tabular}
	\end{center}
	\caption{\label{table:1} Flat priors on the main cosmological parameters constrained in this paper.}
\end{table}

\subsection{Datasets} \label{sec:2.2}
\textbf{CMB: Planck Public Release (PR) 4}: We make use of the most up-to-date measurements of the Cosmic Microwave Background (CMB) temperature and E-mode polarization power spectra on both large angular scales (low-$\ell$) and small angular scales (high-$\ell$) from the \textit{Planck} satellite. For the high-$\ell$ ($30 < \ell < 2500$) TT, TE, and EE spectra, we adopt the latest HiLLiPoP likelihoods, as presented in \cite{Tristram:2023haj}. The low-$\ell$ ($\ell < 30$) EE data are analyzed using the most recent LoLLiPoP likelihoods, also described in \cite{Tristram:2023haj}. Both HiLLiPoP and LoLLiPoP are based on the \textit{Planck} Public Release (PR) 4, which corresponds to the latest reprocessing of LFI and HFI observations with the unified NPIPE pipeline. This reprocessing provides an expanded dataset with reduced noise levels and improved cross-frequency consistency \cite{Planck:2020olo}. For the low-$\ell$ TT data, we employ the Commander likelihood from the \textit{Planck} 2018 release \cite{Planck:2018vyg}. Throughout this work, we collectively denote this combination of likelihoods as \textbf{``Planck PR4."}

\textbf{CMB lensing: Planck PR4+ACT DR6}. CMB experiments also provide measurements of the gravitational lensing potential power spectrum, $C_\ell^{\phi \phi}$, obtained via 4-point functions. In this work, we employ the most recent NPIPE PR4 \textit{Planck} CMB lensing reconstruction \cite{Carron:2022eyg}, together with the Data Release 6 (DR6) from the Atacama Cosmology Telescope (ACT), version 1.2 \cite{ACT:2023kun,ACT:2023ubw}. 
As recommended by the ACT collaboration, 
we adopt the high-precision likelihood settings \cite{ACT:2023kun}. For brevity, we refer to this combined dataset as \textbf{``lensing"}.  
We also refer to the \textbf{Planck PR4+lensing} combination simply as \textbf{``CMB.''}

\textbf{BAO: DESI Data Release (DR) 2}.We include the most recent Baryon Acoustic Oscillation (BAO) measurements from Data Release 2 of the Dark Energy Spectroscopic Instrument (DESI) collaboration \cite{DESI:2025zgx} (for comparison with the earlier DR1 results, see \cite{DESI:2024mwx}). The DR2 dataset covers multiple tracers across a wide redshift range: the Bright Galaxy Sample (BGS, $0.1 < z < 0.4$), the Luminous Red Galaxy Sample (LRG, $0.4 < z < 0.6$ and $0.6 < z < 0.8$), the Emission Line Galaxy Sample (ELG, $1.1 < z < 1.6$), the combined LRG and ELG sample in the overlapping range (LRG+ELG, $0.8 < z < 1.1$), the Quasar Sample (QSO, $0.8 < z < 2.1$), and the Lyman-$\alpha$ Forest Sample (Ly$\alpha$, $1.77 < z < 4.16$). For brevity, we collectively denote this dataset as \textbf{``DESI."}  Since we are not using any other BAO data, we also often refer to this dataset simply as \textbf{``BAO''}.

\textbf{SNeIa: Pantheon+}.We utilize the most recent Type Ia Supernova (SNeIa) luminosity distance measurements from the Pantheon+ compilation \cite{Scolnic:2021amr}, comprising 1550 spectroscopically confirmed SNeIa over the redshift range $0.001 < z < 2.26$. For our analysis, we adopt the publicly available likelihood from \cite{Brout:2022vxf}, which incorporates both statistical and systematic covariance. To reduce the impact of peculiar velocities on the Hubble diagram, this likelihood imposes a cut at $z > 0.01$. Throughout this work, we denote this dataset as \textbf{``PAN+"}.

\textbf{SNeIa: DES Year 5}. We employ the luminosity distance measurements from the most recent supernova sample released by the Dark Energy Survey (DES) as part of their Year 5 data release \cite{DES:2024tys}. This catalog consists of 1635 photometrically classified SNeIa spanning the redshift range $0.1 < z < 1.13$. We denote this dataset as \textbf{``DESY5"}.

We note that the PAN+ and DESY5 samples have overlapping supernovae. To avoid double counting, these two datasets are never used together in any single analysis.  Also, we do not include the Union3 SNe~Ia sample \citep{Rubin:2023jdq}, as the DESI Collaboration has shown that it yields parameter constraints and results intermediate between those obtained from the Pantheon+ and DES~Year~5 samples. Thus, employing the Pantheon+ and DES~Year~5 samples provides the two extreme cases of interest, and that is sufficient for drawing any conclusions regarding cosmological parameters; and we save valuable computational resources by omitting Union3.

\textbf{Weak Lensing: DES Year 1.} We include the likelihood from the joint analysis of galaxy clustering and weak gravitational lensing based on 1321~deg$^2$ of \textit{griz} imaging data from the first year of the Dark Energy Survey \citep{DES:2017myr}. We refer to this dataset as \textbf{``WL"} throughout.


\section{Numerical results} \label{sec:3}
\begin{table*}
	\begin{center}
	\resizebox{\textwidth}{!}{
		\begin{tabular}{cccccccc}
			\toprule
			\toprule
 			\vspace{0.cm}
			Parameter  &    Planck PR4  &    Planck PR4  &    Planck PR4  &    Planck PR4  &    Planck PR4  & Planck PR4   \\\vspace{0.cm}
			 &  + lensing +DESI  &  +lensing  + DESI &    +lensing + DESI & + lensing + DESI&  + lensing+ DESI &  + lensing+ DESI  \\\vspace{0.2cm}
			 &                               &  + PAN+                   &    + DESY5              &  + WL                   &   + PAN+ + WL      &   + DESY5 + WL  \\
			\midrule
			\hspace{1mm}
			\vspace{ 0.2cm}
			$\Omega_b h^2$   &    $0.02263\pm0.00018$  &    $0.02264\pm0.00017$  &    $0.02266\pm0.00017$ & $0.02266\pm0.00018$ & $0.02265\pm0.00018$ & $0.02268\pm0.00017$  \\ \vspace{ 0.2cm}
			
			$\Omega_c h^2$  &    $0.1197\pm0.0028$  &    $0.1197\pm0.0028$  &    $0.1198\pm0.0028$ & $0.1187\pm0.0028$ & $0.1187\pm0.0027$ & $0.1188\pm0.0027$  \\ \vspace{ 0.2cm}
			
			$\tau$  &    $0.0590\pm0.0066$  &    $0.0591\pm0.0066$  &      $0.0591 \pm 0.0066$ & $0.0583 \pm 0.0065$ & $0.0584 \pm 0.0065$ & $0.0585 \pm 0.0065$   \\ \vspace{ 0.2cm}
			
			$n_s$  &    $0.983\pm0.008$  &     $0.983\pm0.008$ &   $0.984\pm0.008$ & $0.983\pm0.008$ & $0.983\pm0.008$ & $0.984\pm0.008$ \\ \vspace{ 0.2cm}
			
			${\rm{ln}}(10^{10} A_s)$  &    $3.044\pm0.016$  &   $3.044\pm0.016$   &     $3.044\pm0.016$ & $3.040\pm0.016$ & $3.044\pm0.016$ &  $3.044\pm0.015$ \\ \vspace{ 0.2cm}
			
		100$\Theta_s^{*}$  &    $1.04077\pm0.00038$   &     $1.04077\pm0.00037$  &  $1.04077\pm0.00038$ & $1.04085\pm0.00038$ & $1.04085\pm0.00038$ & $1.04085\pm0.00038$ \\ \vspace{ 0.2cm}

			$\sum m_\nu$ (eV)  &    $<0.161$ (2$\sigma$)  &    $<0.163$ (2$\sigma$)  &    $<0.153$ (2$\sigma$) & $<0.199$ (2$\sigma$) & $<0.196$ (2$\sigma$) &  $<0.186$ (2$\sigma$) \\ \vspace{ 0.2cm}
			
			$N_{\textrm{eff}}$  &    $3.26\pm0.18$  &    $3.27\pm0.18$  &    $3.28\pm0.18$ & $3.24\pm0.18$ & $3.24\pm0.18$ & $3.25\pm0.18$ \\ \vspace{ 0.2cm}
			
			$w_0$  &    $<-0.874$  &    $-0.946^{+0.025}_{-0.037}$ &    $-0.910\pm0.033$ &  $<-0.877$ & $-0.946^{+0.025}_{-0.037}$ &  $-0.910\pm0.033$ \\ \vspace{ 0.2cm}
			
			$w_a$  &    $0.007^{+0.058}_{-0.073}$  &    $0.006^{+0.043}_{-0.079}$  &    $-0.039^{+0.039}_{-0.075}$ & $0.011^{+0.057}_{-0.074}$ & $0.006^{+0.045}_{-0.078}$  &  $-0.039^{+0.039}_{-0.075}$ \\ \vspace{ 0.2cm}

			$\alpha_s$  &     $0.0008\pm0.0071$ &   $0.0009\pm0.0071$  &    $0.0013\pm0.0071$ & $0.0020\pm0.0071$ & $0.0018\pm0.0071$ & $0.0022\pm0.0071$\\ 
			
				\multirow{3}{*}{$A_{\textrm{lens}}$} &    $1.092^{+0.039}_{-0.045}$ (1$\sigma$),   &    $1.093^{+0.038}_{-0.046}$ (1$\sigma$),   &    $1.098^{+0.038}_{-0.045}$ (1$\sigma$), & $1.112^{+0.038}_{-0.045}$ (1$\sigma$), & $1.111^{+0.037}_{-0.044}$ (1$\sigma$) & $1.113^{+0.037}_{-0.043}$ (1$\sigma$),  \\ \vspace{ 0.2cm} 
			&  $1.092^{+0.090}_{-0.080}$ (2$\sigma$), &  $1.093^{+0.085}_{-0.082}$ (2$\sigma$), & $1.098^{+0.087}_{-0.078}$ (2$\sigma$), & $1.112^{+0.087}_{-0.077}$ (2$\sigma$), & $1.111^{+0.086}_{-0.076}$ (2$\sigma$), & $1.113^{+0.085}_{-0.075}$ (2$\sigma$),
				\\ \vspace{ 0.2cm} 
                & $1.092^{+0.12}_{-0.10}$ (3$\sigma$) & $1.093^{+0.12}_{-0.10}$ (3$\sigma$) & $1.098^{+0.12}_{-0.10}$ (3$\sigma$) & $1.112^{+0.12}_{-0.098}$ (3$\sigma$) & $1.111^{+0.12}_{-0.098}$ (3$\sigma$) & $1.113^{+0.11}_{-0.097}$ (3$\sigma$)\\ 
                
                \midrule

		   	$H_0$ (km/s/Mpc)  &    $68.5\pm1.1$  &     $68.44\pm0.96$    &    $67.91\pm0.96$ & $68.5\pm1.1$ & $68.32\pm0.98$ &$67.79\pm0.97$   \\ \vspace{ 0.2cm}
		   
		   $S_8$  &    $0.799^{+0.015}_{-0.012}$  &    $0.799^{+0.015}_{-0.012}$  &    $0.800^{+0.015}_{-0.012}$ & $0.790^{+0.014}_{-0.012}$ & $0.790^{+0.014}_{-0.011}$  & $0.791^{+0.014}_{-0.011}$  \\ 
			
			\bottomrule
			\bottomrule

	\end{tabular}
}
	\end{center}
	\caption{\label{table:2}\footnotesize Bounds on cosmological parameters in the 12 parameter extended model. Unless otherwise specified, marginalized limits are given at 68\% C.L. whereas upper limits are given at 95\% C.L. Note that $H_0$ and $S_8$ are derived parameters.}
\end{table*}

Here we present constraints on the 12-parameter spatially flat ($\Omega_k=0$) cosmological model, with emphasis on the lensing amplitude parameter, $A_{\rm lens}$. Table~\ref{table:2} provides the full set of marginalized parameter bounds for various combinations of Planck PR4, DESI BAO, supernovae (PAN+ and DESY5), and DES Year 1 weak lensing data.

%

\begin{figure}[tbp]
\begin{center}
\includegraphics[width=.99\linewidth]{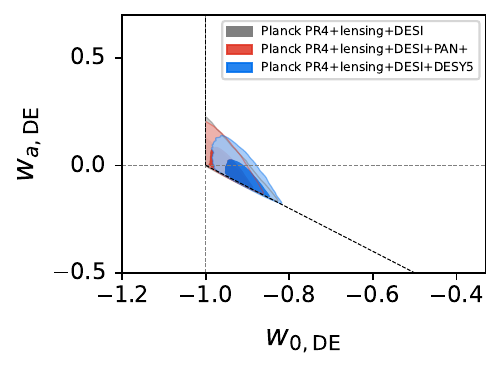}
\caption{\label{fig:1} 68\% and 95\% marginalised  contours in the $w_{0,\rm DE} - w_{a,\rm DE}$ plane for different data combinations. The area to the right of the vertical dashed blue line and above the slanted dashed blue line represents the parameter space corresponding to quintessence-like or non-phantom dark energy.}
\end{center}
\end{figure}

\subsection{$w_{0,\rm DE}$ and  $w_{a,\rm DE}$}
As shown in Figure~\ref{fig:1}, when the parameter space is restricted to non-phantom dark energy ($w(z)\geq -1$), the combination of CMB (Planck PR+lensing) and BAO (DESI) data is still consistent with the cosmological constant ($w_{0,\rm DE}=-1$, $w_{a,\rm DE}=0$) within $1\sigma$. The inclusion of Pantheon+ excludes the cosmological constant at $1\sigma$, but it remains allowed within $2\sigma$. In contrast, adding the DES Year~5 supernova dataset to the CMB+BAO combination rules out the cosmological constant at more than $2\sigma$. Therefore, the DESY5 supernova data appear to favor a dynamical dark energy scenario even under the restriction to the non-phantom region of the dark energy equation of state. Addition of WL data has negligible impact on the DE EoS parameter constraints.

\subsection{$A_{\rm lens}$}

From Table~\ref{table:2}, it is clear that the CMB+DESI and the CMB+DESI+SNe (for both PAN+ and DESY5) combinations lead to a significant 2$\sigma$+ lensing anomaly (i.e. departure from $A_{\rm lens} = 1$). This is in stark contrast to the Planck PR4 Hillipop result in $\Lambda$CDM, where the $A_{\rm lens}$-anomaly was reduced to $<1\sigma$ \cite{Tristram:2023haj}. Our result shows for the first time that physically motivated models like Quintessence can lead to a 2$\sigma$+ lensing anomaly even with Planck PR4 likelihoods. This can be explained by the correlation between $A_{\rm lens}$ and the DE EoS parameters as recently studied in \cite{Park:2025azv,Park:2024pew,RoyChoudhury:2025dhe}.

\begin{figure*}[tbp]
\begin{center}
\includegraphics[width=.49\linewidth]{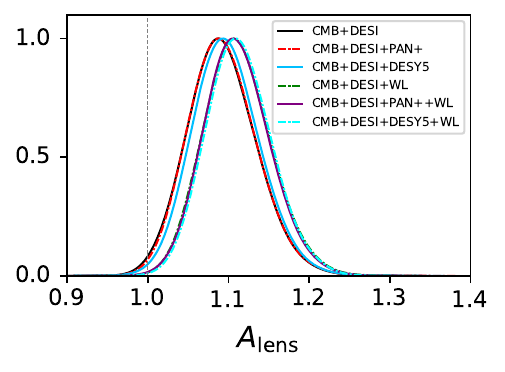}
	\hfill
\includegraphics[width=.49\linewidth]{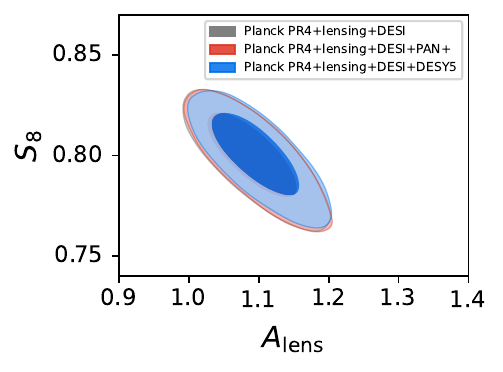}
\caption{\label{fig:2} The left panel shows the 1D posterior distributions of $A_{\rm lens}$ for various data combinations. The right panel shows its 2D correlation plots with the $S_8$ parameter. We note that dataset combinations with WL included leads to a 3$\sigma$+ lensing anomaly due to the strong negative correlation with $S_8$. Even without the WL data, the lensing anomaly is 2$\sigma$+.}
\end{center}
\end{figure*}

The right panel of Figure~\ref{fig:2} demonstrates a pronounced negative correlation between $S_8$ and $A_{\rm lens}$. WL data prefers slightly lower value of $S_8$ than the CMB+BAO+SNe combination. From the left panel, it is evident that the inclusion of WL data shifts the preferred values of $A_{\rm lens}$ towards even higher values, as expected from this correlation. As reported in Table~\ref{table:2}, once WL data are incorporated, the inferred $A_{\rm lens}$ departs from unity at a significance greater than $3\sigma$, which is a significant result.
Although residual systematics in the weak lensing reconstruction may play a role, the anomaly could equally well indicate new physics influencing the growth of structure or the lensing potential, particularly if the data are interpreted at face value. An important implication is that if the underlying cosmological model corresponds to a QFT-realistic scenario such as quintessence, it must also account for the presence of a significant lensing anomaly and explain its origin.

%

We note that the DES Year~1 WL data used in this analysis yields $S_8 = 0.783^{+0.021}_{-0.025}$ within the $\Lambda$CDM framework \citep{DES:2017myr}. This value differs from the $S_8$ inferred in the 12-parameter model with CMB+BAO and CMB+BAO+SNe by only $\sim0.5$--$0.6\sigma$, as can be read from Table~\ref{table:2}. Hence, no significant $S_8$ tension is present, and the WL data can be consistently combined with the CMB+BAO or CMB+BAO+SNe datasets considered in this work.  We further note that cosmological probes independent of both CMB and weak lensing also favor slightly lower values of $S_8$ compared to the CMB (see, e.g., \cite{Silva:2025twg}, which combines DESI DR2 BAO with the SDSS BOSS DR12 full-shape power spectrum). This suggests that the $S_8$ values inferred from weak lensing measurements may indeed be quite plausible.

Also, as noted previously in \cite{RoyChoudhury:2024wri}, the $S_8$ tension with DES Year~3 data is only at the level of $1.4\sigma$ (see also \cite{Tristram:2023haj}), with DES~Y3 preferring slightly lower values of $S_8$ compared to DES~Y1 \citep{DES:2021wwk}. Consequently, had we adopted the DES~Y3 likelihoods, we would likely have found stronger evidence for the lensing anomaly. It is important to remark, however, that recent results from the completed KiDS survey yield somewhat higher values of $S_8$, reporting $S_8 = 0.815^{+0.016}_{-0.021}$ within the $\Lambda$CDM framework \citep{Wright:2025xka}. Thus, it remains unclear whether a $>3\sigma$ lensing anomaly would persist when employing the KiDS dataset. Nevertheless, caution is necessary, since the model favored by the DESY5 SN data departs significantly from the standard $\Lambda$CDM scenario, and thus the $S_8$ constraints inferred from KiDS may also shift accordingly. And even without the WL data, our results are still significant, i.e., a 2$\sigma$+ lensing anomaly with CMB+BAO+SNe.

%
\begin{figure*}[tbp]
\begin{center}
	\includegraphics[width=.32\linewidth]{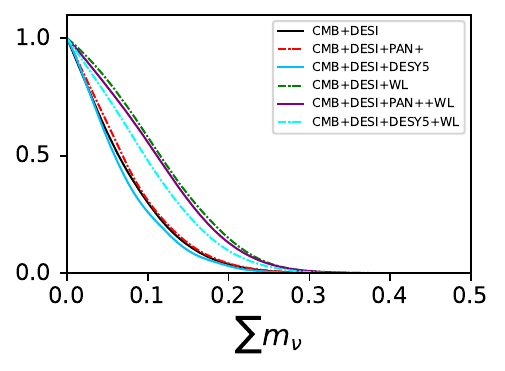}
	\includegraphics[width=.32\linewidth]{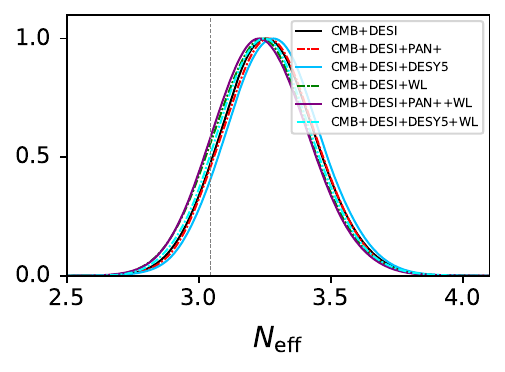}
	\includegraphics[width=.32\linewidth]{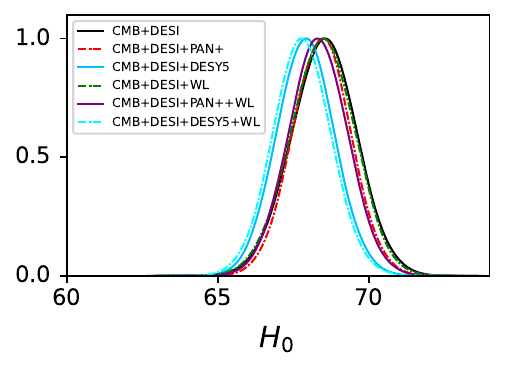}
\caption{\label{fig:3} The left, middle, and right panels show the 1D posterior distributions of $\sum m_{\nu}$, $N_{\rm eff}$, and $H_0$ (km/s/Mpc) respectively, for various data combinations. The dashed vertical line in the middle panel corresponds to the standard model value of $N_{\rm eff} = 3.044$.}
\end{center}
\end{figure*}

\subsection{$\sum m_{\nu}$, $N_{\rm eff}$, and $H_0$}

Figure \ref{fig:3} shows the probability posterior distributions of $\sum m_{\nu}$, $N_{\rm eff}$, and $H_0$. We find that the bounds on $\sum m_{\nu}$ are tighter than in the case where the full range of DE EoS parameters was allowed, as shown in our recent study \cite{RoyChoudhury:2025dhe}. This behavior is expected from the well-known geometric degeneracy between the DE EoS and $\sum m_{\nu}$, first noted in \cite{Hannestad:2005gj}. In particular, models with $w(z)<-1$ tend to prefer larger neutrino masses, while scenarios with $w(z)>-1$ can yield bounds stronger than those obtained in $\Lambda$CDM \cite{Vagnozzi:2018jhn,RoyChoudhury:2018vnm}. As illustrated in Figure~3 of \cite{RoyChoudhury:2025dhe}, the $w_a$ parameter is strongly correlated with $\sum m_{\nu}$. In the present analysis, however, we do not obtain bounds stronger than $\Lambda$CDM, since we also vary $A_{\rm lens}$, which is itself strongly positively correlated with $\sum m_{\nu}$ \cite{RoyChoudhury:2019hls}; higher $A_{\rm lens}$ values therefore drive larger neutrino mass estimates. Moreover, the inclusion of WL data slightly relaxes the bounds on $\sum m_{\nu}$, reflecting the strong negative correlation between $S_8$ and $\sum m_{\nu}$: larger neutrino masses enhance the suppression of the small-scale matter power spectrum, thereby lowering $S_8$.

We find that all dataset combinations yield slightly higher values of $N_{\rm eff}$, excluding the standard model prediction of $N_{\rm eff}=3.044$ at more than $1\sigma$ but less than $2\sigma$, indicating no statistically significant deviation. We also observe that the Hubble tension persists at the $\sim 3.0$--$3.7\sigma$ level, depending on the dataset combination. At the two extreme ends, we find $H_0 = 68.5 \pm 1.1$ km/s/Mpc with CMB+DESI+WL, and $H_0 = 67.79\pm 0.97$ km/s/Mpc with CMB+DESI+DESY5+WL. Hence, the tension remains robust, and the extensions explored in this work are insufficient to resolve it, pointing instead toward the need for new physics.

We note that it is well established that $N_{\rm eff}$ and $H_0$ are positively correlated (see, for instance, the rightmost panel of Figure~6 in \cite{RoyChoudhury:2025dhe}). As discussed in the Introduction, an increase in $N_{\rm eff}$ enhances the radiation energy density in the early Universe, leading to a faster pre-recombination expansion rate and consequently a higher value of $H_0$. Interestingly, within the NPDDE framework, we obtain slightly larger values of $N_{\rm eff}$—and correspondingly higher $H_0$—than in the dynamical dark energy scenario with the full allowed range of $(w_0, w_a)$ (see Table~\ref{table:2} and compare with Table~2 of \cite{RoyChoudhury:2025dhe}). This outcome is somewhat non-trivial, given that phantom dark energy is generally expected to drive a higher expansion rate and thus a larger $H_0$.

Also, a higher value of $N_{\rm eff}$ leads to an enhanced suppression of the CMB temperature and polarization power spectra \citep{Baumann:2018muz,Baumann:2019keh}. This effect can be partially compensated by increasing $n_s$ and $\alpha_s$, resulting in a positive correlation of $N_{\rm eff}$ with both parameters. Consequently, we obtain slightly higher values of $n_s$ and $\alpha_s$ compared to those reported in \citet{RoyChoudhury:2025dhe}, where the full parameter range of dynamical dark energy was allowed. This behavior may have implications for the viability of various inflationary models \citep{Gerbino:2016sgw,RoyChoudhury:2022rva}. However, we find that $N_{\rm eff}$ and $\alpha_s$ are not correlated with $A_{\rm lens}$; therefore, our main conclusion regarding the lensing anomaly remains unaffected by variations in these parameters.


\section{Conclusions} \label{sec:4}

In this work, we have investigated an extended 12-parameter cosmological framework that incorporates non-phantom dynamical dark energy (NPDDE) in a spatially flat cosmology. By construction, this framework enforces the physically motivated condition that the dark energy equation of state (EoS) remains non-phantom at all times, i.e., $w(z) \geq -1$ for the entire cosmic history. This restriction ensures consistency with the class of quintessence-like models based on single scalar-field dynamics, thereby avoiding theoretical pathologies commonly associated with phantom scenarios. The parameter space we consider extends the standard six parameters of the $\Lambda$CDM model by including the dynamical dark energy parameters $(w_0, w_a)$, the scaling of the CMB lensing amplitude $(A_{\rm lens})$, the sum of neutrino masses $(\sum m_\nu)$, the effective number of relativistic species $(N_{\rm eff})$, and the running of the scalar spectral index $(\alpha_s)$.  

To constrain this model, we employed a comprehensive combination of state-of-the-art cosmological datasets. In particular, we used Baryon Acoustic Oscillation (BAO) measurements from the Dark Energy Spectroscopic Instrument (DESI) Data Release 2, CMB temperature and polarization spectra from Planck Public Release (PR) 4, CMB lensing from both Planck PR4 and Atacama Cosmology Telescope Data Release 6, uncalibrated Type Ia supernovae (SNe) data from the Pantheon+ and Dark Energy Survey Year 5 (DESY5) samples, and weak lensing (WL) data from DES Year 1. This joint analysis leverages the complementary constraining power of geometrical probes, cosmic shear measurements, and high-precision CMB data, thereby enabling us to test the viability of NPDDE models within a significantly enlarged cosmological parameter space.  

Our key result is the robust detection of a lensing anomaly at greater than $3\sigma$ significance when combining CMB+BAO+WL data, as well as when extending the dataset combination to include SNe. Without the WL data, the significance of the lensing anomaly is still 2$\sigma$+, even with Planck PR4 Hillipop+LoLLipop likelihoods which show agreement with the consistency-check parameter $A_{\rm lens}=1$ at $<1\sigma$ in $\Lambda$CDM. Specifically, we find that
$A_\mathrm{lens}$ deviates from its theoretically expected value of
unity, i.e., $A_\mathrm{lens}> 1$, indicating that the data prefer a
higher level of CMB lensing than predicted by the NPDDE model. This result highlights a persistent tension: while our parameter space explicitly accommodates quintessence-like dark energy, doing so necessarily requires accepting the presence of a non-negligible lensing anomaly.  

The implications of this finding are twofold. First, it strengthens the case that the lensing anomaly in a realistic Quintessence-like dark energy model is not an artifact of specific dataset choices, but instead emerges robustly when diverse probes are combined. Second, it suggests that models of quintessence-like dark energy cannot be evaluated in isolation: any consistent cosmological framework allowing for realistic dynamical dark energy must also contend with the presence of this anomaly. Possible explanations may include unresolved systematic effects in the datasets, unaccounted-for physical processes in the late universe, or extensions to the cosmological model that go beyond the NPDDE parameter space considered here.

Apart from the lensing anomaly, we find that the bounds on $\sum m_\nu$ are tighter than in scenarios where the full dark energy equation-of-state parameter space is allowed, reflecting the well-known degeneracy between neutrino mass and dark energy dynamics. Although models with $w(z)>-1$ can in principle yield bounds stronger than $\Lambda$CDM, our analysis does not achieve this once $A_{\rm lens}$ is varied, as it correlates positively with $\sum m_\nu$. The inclusion of weak lensing data further relaxes the bounds, owing to the negative correlation between $S_8$ and $\sum m_\nu$.  Also, across all dataset combinations, we obtain mildly higher values of $N_{\rm eff}$, excluding the standard prediction at just above $1\sigma$ but below $2\sigma$, without any statistically significant deviation. The Hubble tension remains at the $\sim 3.0$--$3.7\sigma$ level depending on the dataset combination, indicating that the extensions explored here cannot resolve it and pointing instead toward the need for new physics.

Future work will benefit from upcoming high-precision datasets, particularly from CMB experiments such as Simons Observatory \citep{SimonsObservatory:2018koc} and LiteBIRD \citep{LiteBIRD:2023zmo}, further DESI releases, and large-scale galaxy clustering surveys (e.g., Vera Rubin Observatory \citep{LSST:2008ijt} and Euclid \citep{Amendola:2016saw}). These will be crucial in clarifying whether the lensing anomaly persists at high statistical significance and in disentangling its physical origin. In this context, our results emphasize the importance of jointly analyzing dark energy phenomenology and secondary CMB anisotropy effects, since the viability of realistic quintessence models may ultimately hinge on the resolution of the lensing anomaly.

\begin{acknowledgments}
We acknowledge the use of the HPC facility at ASIAA (https://hpc.tiara.sinica.edu.tw/) where the numerical analyses were done. SRC acknowledges support from grant No. I-IAA-ROY. TO acknowledges support from the Taiwan National Science and Technology Council of Taiwan under grants Nos. NSTC 112-2112-M-001-034-, NSTC 113-2112-M-001-011- and NSTC 114-2112-M-001-004-, and the Academia Sinica Investigator Project Grant No. AS-IV-114-M03 for the period of 2025–2029.  KU acknowledges the support from grant Nos. AS-IA-112-M04, NSTC 112-2112-M-001-027-MY3, and I-SuMIRe.
\end{acknowledgments}



\bibliographystyle{apj}
\bibliography{biblio.bib}

\end{document}